\documentclass[prb,reprint]{revtex4-1} 

\usepackage[colorlinks, allcolors=blue]{hyperref}
\usepackage{amsmath}  
\usepackage{amsfonts} 
\usepackage{graphicx} 
\graphicspath{ {images/} }

\begin{document}


\title{Low-cost Quadrature Optical Interferometer}

\author{Tanner M.\ Melody}
\affiliation{Department of Physics and Astronomy, San Jos\'e State University, San Jose, CA 95192, USA}
\author{Krishna H.\ Patel}
\affiliation{Department of Physics and Astronomy, San Jos\'e State University, San Jose, CA 95192, USA}
\author{Peter K.\ Nguyen}
\affiliation{Department of Physics and Astronomy, San Jos\'e State University, San Jose, CA 95192, USA}
\author{Christopher L.\ Smallwood}
\email[Email: ]{christopher.smallwood@sjsu.edu}
\affiliation{Department of Physics and Astronomy, San Jos\'e State University, San Jose, CA 95192, USA}
\date {\today}

\begin{abstract}
We report on the construction and characterization of a low-cost Mach-Zehnder optical interferometer in which quadrature signal detection is achieved by means of polarization control. The device incorporates a generic green laser pointer, home-built photodetectors, 3D-printed optical mounts, a circular polarizer extracted from a pair of 3D movie glasses, and a Python-enabled microcontroller for analog-to-digital data acquisition. Components fit inside of a $12'' \times 6''$ space and can be assembled on a budget of less than US\$500. The device has the potential to make quadrature interferometry accessible and affordable for instructors, students, and enthusiasts alike.
\end{abstract}

\maketitle 

\section{Introduction} 

Optical interference is a phenomenon in which overlapping beams of light act, by means of the superposition principle of electromagnetism, to modulate the energy density that would have otherwise been present in the constituent beams on their own. In turn, the effect results in striking patterns of irradiance fringes---akin to the nodes and antinodes that can be observed in the harmonic modes of a vibrating guitar string---that both verify the wave nature of light and enable precision measurements. Optical interferometers are devices engineered to exploit this phenomenon, and due to (1) the connection between interference fringes and the extremely short length scales associated with optical wavelengths (400--700 nm), (2) the abundance of high-quality manipulation and detection capabilities at these wavelengths, and (3) the invention of lasers, interferometers have long enjoyed wide-ranging relevance in science and technology. Applications include testing fundamental physics principles,\cite{Michelson1887,Abbott2016} measuring velocities and positions,\cite{Bobroff1993,Field2004,Gao2015} characterizing material properties,\cite{Griffiths} and controlling and manipulating both classical and quantum light sources.\cite{Mabrouki1996,Kanseri2008,RodriguezLara2009,Singh2021,Gerry} Beyond this, optical interferometers enjoy a prominent position in classroom physics laboratories, and several different supply companies are currently selling commercialized products and activities.\cite{Thorlabs2022,PASCO2022,TeachSpin2022}

For the most part, interferometers are designed for delicate measurements and/or repeated use, and so construction costs can be large, ranging from about \$3,000 (for an educational apparatus available from Thorlabs),\cite{Thorlabs2022} all the way up to \$1.1 billion [for the laser interferometer gravitational-wave observatory (LIGO)].\cite{LIGOfacts2017} There is a certain utility, however, in exploring the degree to which these instruments can be home-built and/or scaled down to minimal components and cost.\cite{Cave1955,Kim1999,Vollmer2008,Scholl2009,Felekis2010,Oliveira2013,Libbrecht2015,Pathare2016,Muhlberger2021}  Such devices expand the growing body of optics capabilities enabled by the maker movement,\cite{Willis2012,Porter2016,SalazarSerrano2017,Delmans2018,Brekke2020,Mantia2022,DelRosario2022} and could, for example, be assembled by hobbyists or produced at scale and shipped out to students in large-enrollment online classes.

Here we report on the construction of an optical interferometer that can be assembled on a budget of under \$500, and which exhibits both automated data acquisition and quadrature detection capabilities. The interferometer utilizes a green laser pointer, 3D-printed optical mounts, home-built photodetectors, and microcontroller-based analog-to-digital signal conversion. Among the unique aspects of the setup differentiating it from other low-cost interferometers that have been reported to date is the fact that quadrature signals are generated through polarization control, with phase delays between horizontal and vertical polarization components achieved using a filter extracted from a pair of circularly polarized 3D movie glasses. We benchmark interferometer performance by using it to measure the thermal expansion coefficient of an aluminum plate on which the interferometer is mounted. While the results reveal quantitative inaccuracies, qualitative features are robust. We discuss possible sources of error and areas for improvement.

The construction of the instrument described in this manuscript formed the basis of an independent research project conducted by undergraduates and a master's student in our laboratory, and we found it to be an excellent means of teaching students about the real-world applications of interferometers, quadrature detection, wave plates, Jones matrices, circuitry, and experimental control protocols. The device has the potential to be used in the same manner at other institutions, or could perhaps be manufactured at scale and utilized by students in larger and more structured classroom settings.

\section{Theoretical background}
\label{sec:theory}

Like interferometers more generally, quadrature-detected interferometers have long been employed as scientific measurement devices, and they can take on a variety of different geometrical configurations.\cite{Peck1953,Bouricius1970,Barker1972,Elsworth1973,Downs1979,Bobroff1993,Hogenboom1998,Eom2001} Devices of this sort are collectively defined by their ability to generate a pair of output signals that have been engineered to be ``in quadrature" with each other, which is to say that there is a $\pi/2$-radian (or quarter wavelength) phase shift between the way that the two different signals monitor the interference effects of the overlapping beams.

Among the more common means of obtaining quadratures is through polarization control, in which case the optical path difference between the interferometer's arms for vertically polarized light (for example) is phase-shifted by a quarter of a wavelength relative to the optical path difference for horizontally polarized light. Because of the vector nature of electromagnetism, the interference effects associated with these two different polarization states can be independently examined (note that horizontal and vertical polarization states do not interfere with each other, a fact codified in the Fresnel-Arago laws), and the signals can be filtered and mapped onto distinct quadrature-shifted irradiance measurements at the point of a pair of detectors. In turn, polarization control is often achieved by means of optical retarders, and in the implementation described in this work, we have followed this approach by inserting a quarter-wave plate retarder and a few judiciously placed linear polarizers into an otherwise standard Mach-Zehnder configuration.

\begin{figure}[tb]\centering
\includegraphics[width=3.4in]{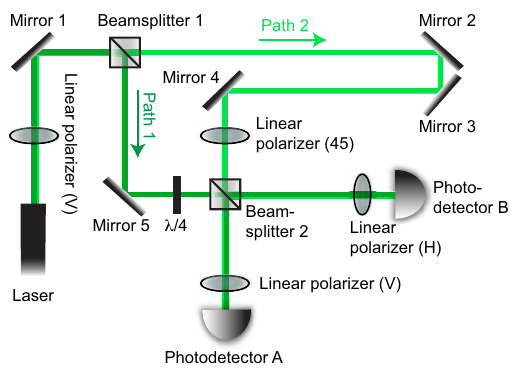}
\caption{\label{schematic}Schematic illustration of an example polarization-based quadrature Mach-Zehnder interferometer.}
\end{figure}

Figure \ref{schematic} shows an example polarization-based quadrature optical interferometer setup. Beamsplitters 1 and 2 are nonpolarizing beamsplitters. Linear polarizers are labeled with vertical (V), horizontal (H), or +45$^\circ$ rotated (45) transmissive axes, with coordinates specified in a lab frame while looking into the beam (i.e., the $x$-axis is horizontal, the $y$-axis is vertical, and the $z$-axis runs parallel to the direction of beam propagation). The quarter wave plate (labeled $\lambda/4$) is oriented such that its fast axis is rotated $-45^\circ$ downward from the horizontal lab frame axis while looking into the beam. Incoming light is vertically polarized.

If we assume, for simplicity's sake, that the light passing through the interferometer is a harmonic traveling plane wave of angular frequency $\omega$, then the dynamics of this setup can be characterized in terms of a Jones matrix analysis,\cite{Jones1941,Pedrotti,Hecht,Brosseau} where the electromagnetic field vector
\begin{equation}\label{eq:1}
{\bf E}(\ell,t) =  \textrm{Re}\left\{ \left( \tilde{E}_{0 x} \hat{\bf x} + \tilde{E}_{0 y} {\hat{\bf y}} \right) e^{i(k_0\ell-\omega t)} \right\}
\end{equation}
is described by a two-component amplitude vector $\tilde{\bf E}_0$ (i.e., a \emph{Jones vector}) 
\begin{equation}
\tilde{\bf E}_0 = \begin{bmatrix}
\Tilde E_{0x} \\
\Tilde E_{0y}
\end{bmatrix} = \begin{bmatrix}
 E_{0x}{e^{i\phi_x}}\\
 E_{0y}{e^{i\phi_y}}
\end{bmatrix}
\end{equation}
such that
\begin{equation}\label{eq:2}
{\bf E}(\ell,t) =  \textrm{Re}\left\{ \tilde{\bf E}_0 e^{i(k_0\ell-\omega t)} \right\}.
\end{equation}
The $x$-direction, or upper entry, within this representation corresponds to the electromagnetic field's horizontal (or $p$-polarized) component, with positive $x$ pointing to the right while staring into the beam. The $y$-direction, or lower entry, corresponds to the field's vertical (or $s$-polarized) component, with positive $y$ always pointing upward in the lab frame. These conventions are consistent with those of textbooks by Pedrotti\cite{Pedrotti} and Hecht.\cite{Hecht} The variable $\ell\equiv\int n(z) \, dz$ corresponds to optical path length with $n(z)$ being the position-dependent refractive index, and the vacuum wavenumber is $k_0 \equiv 2\pi/\lambda_0 = \omega/c$ with $c = \textrm{299,792,458 m/s}$. Symbols with tildes on top explicitly indicate complex-valued quantities.

Having described the electromagnetic field amplitude in terms of a two-component vector of this sort, the actions of different sorts of optical elements can be described in terms of multiplication operations by different kinds of matrices, termed \emph{Jones matrices}. For example, the action of wave plates oriented such that their fast and slow axes are aligned to the Jones vector coordinate axes can be represented by
\begin{equation} 
\begin{bmatrix}
     e^{i{\phi_x}}&0 \\ 
     0&e^{i{\phi_y}}
    \end{bmatrix}, \label{qwp1}
\end{equation}
where ${\phi_x}$ and ${\phi_y}$ are the phase lags induced by the wave plate along the $x$-axis and $y$-axis, respectively.
For a quarter wave plate, ${\phi_y}-{\phi_x}=\pi/2$. Thus, the matrix for a quarter wave-plate with its fast axis oriented along the $x$-direction can be represented as
\begin{equation}
    \begin{bmatrix}
    1&0 \\ 
    0&i
\end{bmatrix}, \label{qwp2} 
\end{equation}
where a factor of $e^{i\phi_x}$ has been pulled out of the expression and dropped because global phase shifts have no effect on the polarization state and can moreover be erased by shifting the origin of the position or time axis in Eq.~(\ref{eq:1}). 

The Jones matrix for a linear polarizer with its transmissive axis oriented along the $x$-direction is
\begin{equation}
\begin{bmatrix}
    1 & 0 \\ 
    0 & 0
\end{bmatrix}. \label{pol}
\end{equation}
The Jones matrix of a mirror is
\begin{equation}
\left[ \begin{array}{rr}
    1&0 \\ 
    0&-1
\end{array}\right],
\end{equation}
where a minus sign between the $x$- and $y$-directions has been introduced to account for the fact that the coordinate system should rotate by 180$^\circ$ about the $y$-axis to keep the reference frame oriented as if looking back toward the source.

Beamsplitters have somewhat more complicated Jones matrix representations than the representations of other types of optical elements because of their doubled input and output ports, but descriptions remain well established.\cite{Siegman,Beyersdorf1999} We employ in this work a description in which the reflection matrix is given by
\begin{equation}
 \begin{bmatrix}
 r_p&0 \\ 
 0&-r_s
\end{bmatrix},  
\end{equation}
and the transmission matrix is given by
\begin{equation}
 \begin{bmatrix}
 it_p&0 \\ 
 0&it_s
\end{bmatrix},
\end{equation}
regardless of incoming light direction. Here $r_\alpha$ and $t_\alpha$ are the electric field reflection and transmission coefficients, and the beamsplitter is assumed to be lossless. Energy conservation is preserved by the relationship $r_\alpha^2 + t_{\alpha}^2 = 1$, and by the factors of $i$ preceding the coefficients $t_\alpha$.

Finally, birefringent optical elements with axes rotated into orientations other than those of the Jones vector coordinates can be described by means of combining matrices like Eqs.~(\ref{qwp1}), (\ref{qwp2}), and (\ref{pol}) with rotation matrices of the form
\begin{equation}
R(\theta) = 
\left[ \begin{array}{rr}
\cos \theta & -\sin \theta \\ 
\sin \theta & \cos \theta
\end{array}\right].
\end{equation}
For example, a quarter wave plate with its fast axis oriented at $-45^\circ$ relative to the to Jones vector $x$-axis can be described by the matrix
\begin{align}
R(-45^\circ)
\begin{bmatrix}
1 & 0 \\ 
0 & i
\end{bmatrix}
R(45^\circ)
=
\frac{e^{i\pi/4}}{\sqrt{2}}&\begin{bmatrix}
1 & i \\ 
i & 1
\end{bmatrix} \nonumber \\
\to 
\frac{1}{\sqrt{2}}&\begin{bmatrix}
1 & i \\ 
i & 1
\end{bmatrix}.
\end{align}
The Jones matrix for linear polarizer with transmission axis rotated at +45${^\circ}$ relative to the Jones vector $x$-axis can be described by the matrix 
\begin{align}
R(45^\circ)
\begin{bmatrix}
1 & 0 \\ 
0 & 0
\end{bmatrix}
R(-45^\circ)
=
\frac{1}{2}&\begin{bmatrix}
1 & 1 \\ 
1 & 1
\end{bmatrix}.
\end{align}

Having laid out these various definitions, we seek to examine the interference properties of signals measured at Photodetectors A and B in Fig.~\ref{schematic} as a function of varied optical path differences between Beam Paths 1 and 2. The physically measurable quantity of interest in these cases is the irradiance $I(\ell,t)$, which is related to the total electric field ${\bf E}(\ell,t)$ according to the equation
\begin{equation}
I(\ell,t) = n \epsilon_0 c \langle E^2(\ell,t) \rangle,
\end{equation}
where $\epsilon_0$ is the dielectric permittivity of vacuum and the angle brackets indicate a time average of the expression over several optical oscillation periods. In the case at hand, the total electric field ${\bf E}(\ell,t)$ corresponds to the vector sum of the two different component vector fields ${\bf E}_1$ and ${\bf E}_2$ propagating through the interferometer along optical paths $\ell_1$ and $\ell_2$. We can rewrite the equation as
\begin{align}
I &= n \epsilon_0 c \langle \left( {\bf E}_1 + {\bf E}_2 \right)^2 \rangle \\
&= n \epsilon_0 c \langle {\bf E}_1^2 + {\bf E}_2^2 + 2 {\bf E}_1 \cdot {\bf E}_2 \rangle \\
&= 
\underbrace{n \epsilon_0 c \langle E_1^2 \rangle}_{I_1}
+ \underbrace{n \epsilon_0 c \langle E_2^2 \rangle}_{I_2}
+ \underbrace{2 n \epsilon_0 c \langle {\bf E}_1 \cdot {\bf E}_2 \rangle}_{I_{12}}, \label{underbraces}
\end{align}
where the terms in Eq.~(\ref{underbraces}) can be separately labeled $I_1$, $I_2$, and $I_{12}$. The quantities $I_1$ and $I_2$ have no dependence on the relative values of $\ell_1$ and $\ell_2$ whereas $I_{12}$ does, and so we see that $I_{12}$ (often explicitly identified as the expression's {\it interference term}\cite{Pedrotti}) is the main quantity of interest. We can write it as
\begin{align}
I_{12} &= 2 n \epsilon_0 c \langle {\bf E}_1 \cdot {\bf E}_2 \rangle \\
&= 2 n \epsilon_0 c \, \textrm{Re}\left\{ \tilde{\bf E}_1^* \cdot \tilde{\bf E}_2 \right\}.
\end{align}

Employing the matrix methods above and examining the irradiance at Photodetector A gives
\begin{align}
    \tilde{\bf E}_1^{(A)} &= 
    \underbrace{\begin{bmatrix} 0&0 \\ 0&1 \end{bmatrix}}_{\textrm{Polarizer}}
    \underbrace{\begin{bmatrix} r_p&0 \\ 0&-r_s \end{bmatrix}}_{\textrm{Beamsplitter 2}}
    \underbrace{{\frac{1}{\sqrt{2}}}
    \begin{bmatrix} 1&i \\ i&1 \end{bmatrix}}_{\lambda/4} \nonumber\\
    & \quad \times 
    \underbrace{\begin{bmatrix} 1&0 \\ 0&-1 \end{bmatrix}}_{\textrm{Mirror 5}}
    \underbrace{\left[ \begin{array}{cc} r_p & 0 \\ 0 & -r_s \end{array} \right]}_{\textrm{Beamsplitter 1}}
    \underbrace{\begin{bmatrix} 0 \\ 1 \end{bmatrix}
    E_0 \, e^{i(k_0 \ell_1 - \omega t)}}_{\textrm{Vertical light}} \\[10pt]
    &= \frac{r_s^2}{\sqrt{2}} \begin{bmatrix} 0 \\ 1 \end{bmatrix}
    E_0 \, e^{i(k_0 \ell_1 - \omega t) - i\pi}
\intertext{and}
    \tilde{\bf E}_2^{(A)} &= 
    \underbrace{\begin{bmatrix} 0&0 \\ 0&1 \end{bmatrix}}_{\textrm{Polarizer}}
    \underbrace{\begin{bmatrix} it_p&0 \\ 0&it_s \end{bmatrix}}_{\textrm{Beamsplitter 2}}
    \underbrace{{\frac{1}{2}}\begin{bmatrix} 1&1 \\ 1&1 \end{bmatrix}}_{\textrm{Polarizer}}\nonumber\\
    & \quad \times 
    \underbrace{\begin{bmatrix} 1&0 \\ 0&-1 \end{bmatrix}^3}_{\textrm{Mirrors 2--4,}}
    \underbrace{\begin{bmatrix} it_p&0 \\ 0&it_s \end{bmatrix}}_{\textrm{Beamsplitter 1}}
    \underbrace{\begin{bmatrix} 0 \\ 1 \end{bmatrix} 
    E_0 \, e^{i(k_0 \ell_2-{\omega}t)}}_{\textrm{Vertical light}} \\[10pt]
    &= \frac{t_s^2}{2} \begin{bmatrix} 0 \\ 1 \end{bmatrix}
    E_0 \, e^{i(k_0 \ell_2 - \omega t)},
\end{align}
leading to
\begin{align}
   {I}_{12}^{(A)} &\propto \frac{r_s^2t_s^2E_0^2}{2{\sqrt{2}}} \, \textrm{Re} \left\{ e^{ik_0(\ell_2-\ell_1)+i{\pi}} \right\}
\intertext{or (adjusting the origin of $\ell_1$ to remove the $\pi$ phase shift)}
    {I}_{12}^{(A)} &\propto \frac{r_s^2t_s^2E_0^2}{2{\sqrt{2}}} \, \textrm{Re} \left\{ e^{ik_0(\ell_2-\ell_1')} \right\}.
   \label{photodetectorA}
\end{align}
Taking a similar approach to determine the irradiance at Photodetector B gives
\begin{align}
    \tilde{\bf E}_1^{(B)} &= 
    \underbrace{\begin{bmatrix} 1&0 \\ 0&0 \end{bmatrix}}_{\textrm{Polarizer}}
    \underbrace{\begin{bmatrix} it_p&0 \\ 0&it_s \end{bmatrix}}_{\textrm{Beamsplitter 2}}
    \underbrace{{\frac{1}{\sqrt{2}}}\begin{bmatrix} 1&i \\ i&1 \end{bmatrix}}_{\lambda/4} \nonumber\\
    & \quad \times 
    \underbrace{\begin{bmatrix} 1&0 \\ 0&-1 \end{bmatrix}}_{\textrm{Mirror 5}}
    \underbrace{\left[ \begin{array}{cc} r_p & 0 \\ 0 & -r_s \end{array} \right]}_{\textrm{Beamsplitter 1}}
    \underbrace{\begin{bmatrix} 0 \\ 1 \end{bmatrix}
    E_0 \, e^{i(k_0 \ell_1 - \omega t)}}_{\textrm{Vertical light}} \\[10 pt]
    &= -\frac{r_st_p}{\sqrt{2}} \begin{bmatrix} 1 \\ 0 \end{bmatrix}
    E_0 \, e^{i(k_0 \ell_1 - \omega t)}
\intertext{and}
    \tilde{\bf E}_2^{(B)} &= 
    \underbrace{\begin{bmatrix} 1&0 \\ 0&0 \end{bmatrix}}_{\textrm{Polarizer}}
    \underbrace{\begin{bmatrix} r_p&0 \\ 0&-r_s   
    \end{bmatrix}}_{\textrm{Beamsplitter 2}}
    \underbrace{{\frac{1}{2}}\begin{bmatrix} 1&1 \\ 1&1 \end{bmatrix}}_{\textrm{Polarizer}}\nonumber\\
    & \quad \times 
    \underbrace{\begin{bmatrix} 1&0 \\ 0&-1 \end{bmatrix}^3}_{\textrm{Mirrors 2--4,}}
    \underbrace{\begin{bmatrix} it_p&0 \\ 0&it_s \end{bmatrix}}_{\textrm{Beamsplitter 1}}
    \underbrace{\begin{bmatrix} 0 \\ 1 \end{bmatrix} 
    E_0 \, e^{i(k_0 \ell_2-{\omega}t)}}_{\textrm{Vertical light}} \\[10 pt]
    &= - \frac{i r_p t_s}{2} \begin{bmatrix} 1 \\ 0 \end{bmatrix}
    E_0 \, e^{i(k_0 \ell_2-{\omega}t)},
    \end{align}
leading to
\begin{align}
    I_{12}^{(B)} &\propto \frac{r_p r_s t_p t_s E_0^2}{2\sqrt{2}} \, \textrm{Re} \left\{ e^{ i k_0 (\ell_2-\ell_1) + i\pi/2 } \right\} \\
    &\propto \frac{r_p r_s t_p t_s E_0^2}{2\sqrt{2}} \, \textrm{Re} \left\{ e^{ i k_0 (\ell_2-\ell_1') - i\pi/2 } \right\}.
    \label{photodetectorB}
\end{align}
By comparing Eqs.~(\ref{photodetectorA}) and (\ref{photodetectorB}), we see that the signals extracted at Photodetectors A and B are 90$^\circ$ out of phase with each other, thereby enabling quadrature detection and analysis.

\begin{figure}[tb]\centering
\includegraphics[width=3.4in]{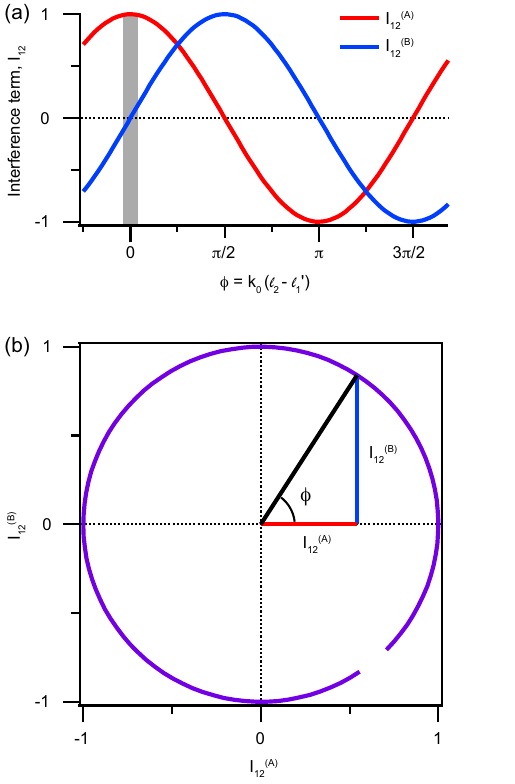}
\caption{\label{theory}Theoretical illustration of the advantages of quadrature detection. 
(a) Illustration of the interference terms $I_{12}^{(A)}$ and $I_{12}^{(A)}$ as defined by Eqs.~(\ref{photodetectorA}) and (\ref{photodetectorB}) as a function of optical phase shift $\phi$.
(b) Lissajous figure plotting $I_{12}^{(B)}$ against $I_{12}^{(A)}$.
}
\end{figure}

Returning back to generalized features, we can see that quadrature detection is useful as a tool for tracking optical path length fluctuations because it helps to remove directional ambiguities. Consider the traces plotted out in Fig.~\ref{theory}(a), showing the dependence of signals $I_{12}^{(A)}$ and $I_{12}^{(B)}$ on the optical path difference $\ell_2-\ell_1'$ in the case where $r_p=r_s=t_p=t_s=1/\sqrt{2}$. For $I_{12}^{(A)}$ or $I_{12}^{(B)}$ considered in isolation, there is an unambiguous mapping from path length difference to irradiance (that is, if we know $\ell_2-\ell_1'$, then we know exactly what value to expect for $I_{12}^{(A)}$). However, the inverse problem is not so well defined, and indeed there are an infinite number of quantities $\ell_2-\ell_1'$ possible if the value of $I_{12}^{(A)}$ is taken as an input. To lift the ambiguity, an optical path difference initial condition can be either independently ascertained or set to zero, and the value of $I_{12}^{(A)}$ can be subsequently tracked as a function of time. There is, however, a remaining problem: when the interference signal comes to an extremum---for example, a maximum, as highlighted in Fig.~\ref{theory}(a) by the vertical gray bar at $\phi=0$---it will in all cases trend back toward equilibrium at times following this, and there is no way to know if that changing signal represents an increase or decrease in the optical path difference. Quadrature detection plugs this hole by monitoring a pair of interference fringes in tandem instead of just a single interference fringe, and when one of the two fringes (e.g., $I_{12}^{(A)}$) comes to an extremum, the other ($I_{12}^{(B)}$) is planted at zero with maximal slope.

Figure \ref{theory}(b) shows an illustration of the $I_{12}^{(A)}$ and $I_{12}^{(B)}$ signals plotted against each other on an $xy$ coordinate scheme (a form of Lissajous figure), and illustrates a convenient graphical means of converting interference fringes back into phase, and by extension, optical path difference. Notice that Eqs.~(\ref{photodetectorA}) and (\ref{photodetectorB}) can be written as
\begin{align}
    I_{12}^{(A)} &\propto E_0^2 \cos \left[ k_0(\ell_2-\ell_1') \right]
\intertext{and}
    I_{12}^{(B)} &\propto E_0^2 \sin \left[ k_0(\ell_2-\ell_1') \right].
\end{align}
Thus, plotting $I_{12}^{(B)}$ on the $y$-axis against $I_{12}^{(A)}$ on the $x$-axis generates values in the $xy$ plane that can be interpreted as forming an angle $\phi = k_0(\ell_2-\ell_1')$ relative to the positive $x$-axis that obeys the relation 
\begin{equation}
    \frac{I_{12}^{(B)}}{I_{12}^{(A)}} \propto \frac{\sin [ k_0(\ell_2-\ell_1')]}{\cos [ k_0(\ell_2-\ell_1']} = \tan [k_0(\ell_2-\ell_1')]. 
\end{equation}
In the case of $r_p=r_s=t_p=t_s=1/\sqrt{2}$, the coefficient of proportionality is 1, and the expression can be inverted to obtain
\begin{equation}
    \phi = k_0(\ell_2-\ell_1') = \textrm{atan2}(I_{12}^{(B)},I_{12}^{(A)}) + 2\pi n,
    \label{phaseextract}
\end{equation}
where $n \in \mathbb{Z}$ and the function $\textrm{atan2}(y,x)$ is the four-quadrant arctangent. The path-length extraction goal is thereby achieved.

\section{Experimental setup}
\label{sec:setup}

\begin{figure}[tb]
    \centering
    \includegraphics[width = 3.4 in]{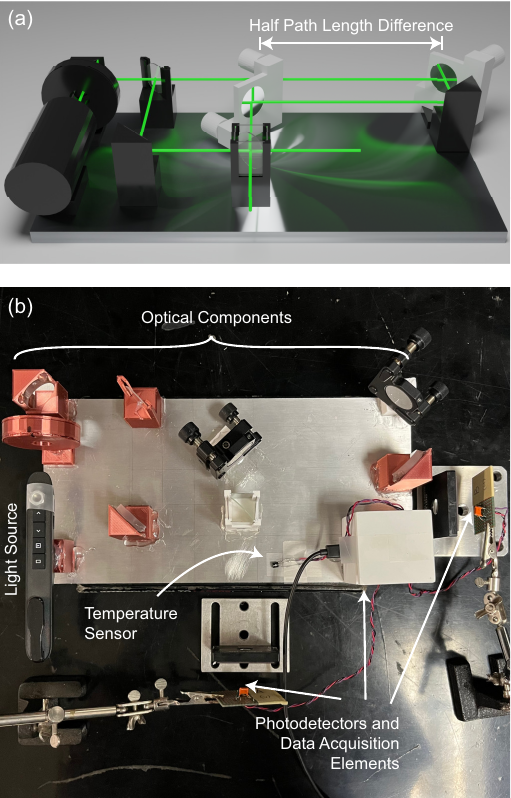}
    \caption{Experimental setup. (a) CAD rendering. (b) Photograph.}
    \label{top_build}
\end{figure}

Having established a theoretical summary of operative principles, we proceed to a description of the physically realized setup. Figure \ref{top_build} shows a 3D CAD rendering and summary photograph of the device, which we constructed according to the Fig.~\ref{schematic} schematic using a $12'' \times 6''$ sheet of 1/4-inch thick aluminum as a base plate. As can be seen in Fig.~\ref{top_build}(a), the system's finite optical path difference between its two interferometer arms gives it the potential of being used to measure the thermal expansion coefficient of the aluminum plate, or alternatively (if the thermal expansion coefficient of aluminum is known) as a high-resolution temperature sensor.

Because the goal of this manuscript is to report on the limits of how inexpensively a polarization-based optical quadrature interferometer can be constructed, it is important to lay out a detailed analysis of components and costs. Table \ref{partslist} summarizes this information in the form of a parts list. The individualized merits of each of these different components are discussed in turn below. 

\begin{table}[tb]
\centering
\caption{Parts list}
\begin{ruledtabular}
\begin{tabular}{l l l l}
Qty & Item Description & Brand/Supplier & Cost \\
\hline	
1 & Green Laser Pointer & Dinofire & \$23.99 \\
1 & Aluminum Plate & Kaylan & \$19.99 \\
1 & Pair of 3D glasses & RealD & \$4.99 \\
1 & Plastic Sheet Polarizer & Izgut & \$12.99 \\
1 & Spool 3D printer filament & Geeetech & \$19.53 \\
2 & Nonpolarizing beamsplitter & Edmund Optics & \$90.00 \\
5 & Silvered Mirrors & Thorlabs & \$164.20\\
2 & Plano-Convex lens & Pre-owned & \$9.00\\
2 & Kinematic mount & Thorlabs & \$79.72 \\
2 & Home-built Photodetectors & Various & \$10.00 \\
1 & LM35 Temperature Sensor & TI & \$2.29 \\
1 & Trinket M0 Microcontroller & Adafruit & \$8.39 \\
1 & Hot Glue Gun & Art Minds & \$13.99\\
1 & Package Hot Glue & Art Minds & \$5.49 \\
1 & Package 5 minute epoxy & Devcon & \$3.69 \\
\hline	
Total & & & \$468.26 \\
\end{tabular}
\end{ruledtabular}
\label{partslist}  
\end{table}

\subsection{Light source}

The typical light source used in commercialized interferometric devices is a helium-neon gas laser, available, for example, from Edmund Optics (Stock \#61-338) for \$1,270. While such lasers offer excellent stability and coherence lengths on the order of 20 cm, this laser cost is far more expensive than the budget we were hoping to achieve. Diode lasers that emit typically in the red spectral range are an inexpensive and ubiquitous alternative, but we chose to avoid these due to concerns about coherence length. Instead, we chose to incorporate a diode-pumped solid-state laser emitting green light at a wavelength of 532 nm for our device (based on frequency-doubling the 1064 nm emission line of Nd$^{3+}$ ions embedded within a host matrix like YAG or YVO$_4$), which combines low cost with long coherence length. Lasers of this sort can be purchased on Amazon for prices ranging from \$20--30. We selected the green-light version of a Dinofire presentation remote for our experiment, purchased online for \$24.99 and pictured on the left side of Fig.~\ref{top_build}(b). Although manufacturer specifications neglected to include information on coherence length, we found the coherence of this item to be nevertheless adequate for our experiment, as demonstrated by the fact that interference fringes between the two arms of the interferometer could be observed at all times when the interferometer was well-aligned. The laser exhibited a significant drop-off in output irradiance over the course of the first several minutes after being turned on, possibly due to a dependence on battery charge level, and so we found it best to wait a minimum of 25 minutes before beginning to collect data.

Not all 532-nm green laser pointers work for the application we had in mind. Aside from the laser used in the demonstrated device, we attempted to use a Pinty 532-nm green laser in the setup. Interference fringes could not be observed, possibly due to the fact that the Pinty laser was designed in such a way that it blinked with a cycling time of about 190 Hz as verified using an oscilloscope.

\subsection{Optical mounts and components}

\begin{figure}[bt]\centering
\includegraphics[width=3.4in]{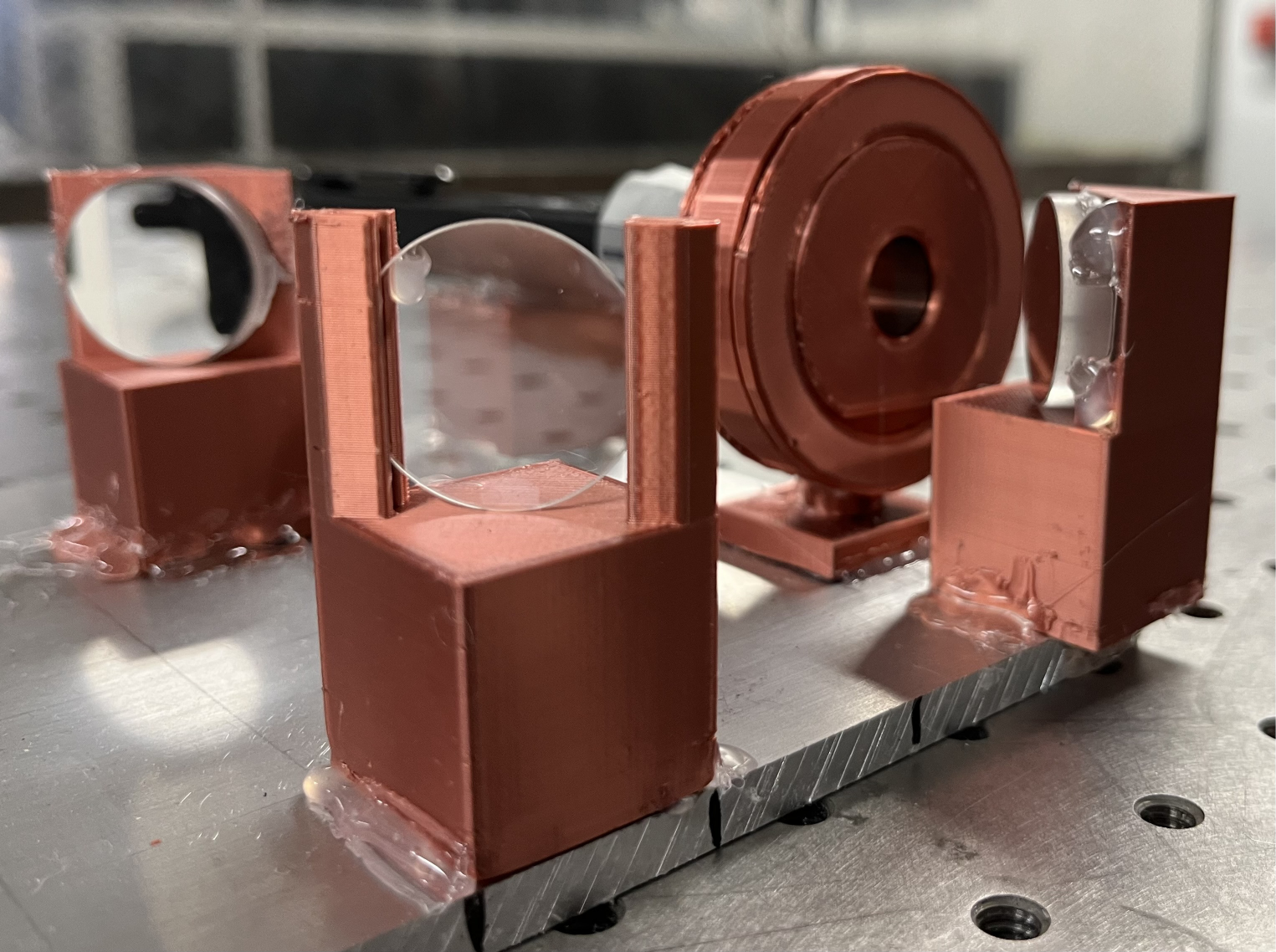}
\caption{Close-up photograph of some of the system's 3D-printed mounts and optics.
}
{\label{opticalmounts}}
\end{figure}

When performing optical experiments, conventional optical mounts are made of metal, with each component (excluding mirrors) costing between \$20-150. One of the simplifications that we made to our own interferometer was to make 3D printed optical mounts from polylactic acid (PLA) plastic filament as illustrated in Fig.~\ref{opticalmounts}. The mirror mounts were printed to position the optics at fixed 45$^{\circ}$ angles with respect to the mirror-mount bases for ease of alignment. The azimuthal rotating mount was printed to be freely adjustable and allowed a linear polarizer to be rotated in place to clean up the laser pointer's initial polarization and direct it to vertical orientation. Each of these mounts costs around \$0.10 to make. We have included links to the OBJ files for all 3D printed optical components in a Github link at the end of the paper. 

Mounts were secured to optical elements and the aluminum base plate using a combination of hot glue and 5-minute epoxy. Although we found 3D-printed parts to be sufficient for fixed-mount components of the system, a pair of kinematic mirror mounts are still required in order to align the two interferometer arms. We used Thorlabs KM100 kinematic mirror mounts to achieve this.

In terms of actual interferometer optical elements, a significant portion of the cost in a traditional polarization-based quadrature interferometer system comes from the polarization optics, specifically linear polarizers (typically ranging from tens to hundreds of dollars per item) and---more critically---wave plate retarders (often priced higher than \$250 per item). In our setup, we procured low-cost versions of linear polarizers by purchasing a sheet of polarizing plastic from the company Izgut (Model \#4335030066). A low-cost quarter wave plate was obtained by installing the lens of a circular polarizer from a pair of 3D movie glasses (RealD) into the setup backwards. Circular polarizers of this type consist of linear polarizers and quarter wave plates stacked on top of each other. If the optic is arranged such that linearly polarized light impinges upon the polarizer before seeing the wave plate, then the polarizer's only impact is to (possibly) reduce the output beam's irradiance, and the light that emerges will be otherwise circularly polarized. Alternate wave plate solutions exist apart from the implementation we have incorporated into this work (see, for example, Edmond Optics $\lambda/4$ Retarder Film, Stock \#14-723), and indeed it has been reported that quarter wave plates can be constructed by means as simple as folding a sheet of clear plastic wrap around a microscope slide.\cite{Hecht} We opted against this final option because of concerns about wave fronts and scattering.

The mirrors we used are manufactured by Thorlabs (Model PF10-03-P01) and cost \$53 each. We attempted to buy cheaper mirrors for a cost of about \$0.05 each (such mirrors can be purchased for example, as craft supplies). However, we found that cheaper mirrors yielded imperfect reflections and corrupted the beam wave fronts. Similarly, we attempted to construct low-cost beamsplitters by attaching one-way window film to microscope slides, but we were unsuccessful in this attempt. The beamsplitters that we ultimately incorporated into the setup were plate beamsplitters purchased from Edmund Optics (Stock \#43-736).

\subsection{Photodetectors and data acquisition elements}

\begin{figure}[tb]\centering
\includegraphics[width=3.4in]{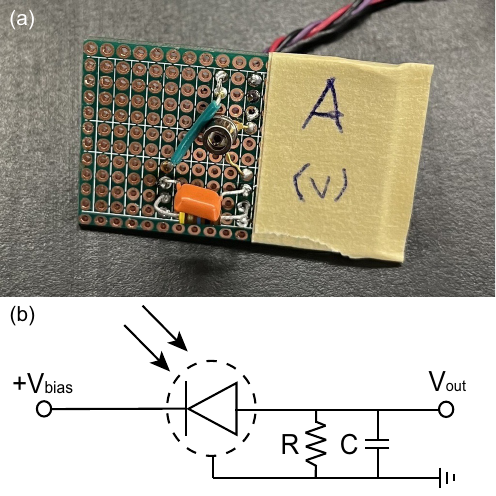}
\caption{Photodetector design elements.
(a) Photograph of Photodetector A.
(b) Associated circuit diagram. A bias voltage was set to 3.3 V. Resistance and capacitance values were $R=10$ M$\Omega$ and $C=10$ nF. 
}
{\label{HomemadePhotodetector}}
\end{figure}

Automated data acquisition capabilities form a critical aspect of nearly all modern optical applications, yet this capability is often found absent in low-cost interferometer reports. In order to facilitate such automated data acquisition, it is necessary to incorporate photodetectors into the setup. Commercially available units are often priced in the range of hundreds of dollars, but we found that a detector consisting of a Hamamatsu S5971 photodiode wired up to a resistor and capacitor arranged in parallel as illustrated in Fig.~\ref{HomemadePhotodetector}  was enough to suit our purposes. We chose a resistance value 10 M$\Omega$ and a capacitance value of 10 nF in our detector design so as to optimize gain and filter out high-frequency noise. 

Because the S5971 is a small-area photodiode, we found it helpful to focus down the light emerging from the interferometer onto the diode active areas using some converging lenses (18-mm focal length) that we had on hand in our laboratory. Such lenses could have alternatively been purchased from a company like Surplus Shed for \$4.50 apiece. Aside from these two focusing lenses used after the beams are combined, there are no other lenses involved in the setup, and so wavefronts are flat at the point where beams are combined, leading to an absence of transverse fringes when the experiment is well-aligned. Because of this, we experienced no loss of precision due to the presence or absence of focusing before the detector.

Photodetectors were held in place by means of alligator clips and soldering stands, which was a design choice dictated by time constraints of the graduation dates of the undergraduate and master's student authors spearheading the project. In future iterations of the device, it would make sense for these mounts to be replaced by 3D-printed mounts affixed to the aluminum plate itself. Because interferometer beams are recombined at the point of the second beam splitter, the distance between this beam splitter and the detectors, and also the materials over which the beams travel while traversing this distance, are irrelevant to interferometer performance.

Although not technically part of interferometer functionality, we needed an independent measurement of temperature in order to conduct the interferometer functionality tests described later on in Sec.~\ref{sec:results}. To achieve this measurement, we incorporated an LM35 temperature sensor manufactured by Texas Instruments as pictured in the center of Fig.~\ref{top_build}(b). The sensor generates a voltage proportional to the ambient temperature in Celsius with a conversion factor of ${1}^{\circ}$C / 10 mV.

\begin{figure}[tb]\centering
\includegraphics[width=3.4in]{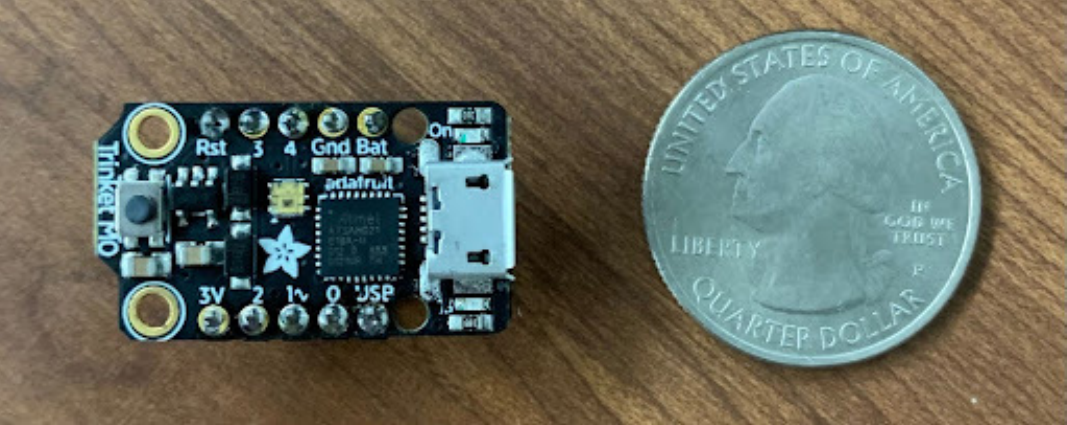}
\caption{Adafruit Trinket M0 microcontroller used for analog-to-digital signal conversion.}
{\label{trinket}}
\end{figure}

Finally, a low-cost analog-to-digital conversion protocol was achieved by means of the Trinket M0 microcontroller purchased from Adafruit, pictured in Fig.~\ref{trinket}. This microcontroller can be coded using CircuitPython, a Python variation that has been specifically designed for microcontroller devices. We programmed the Trinket to have three analog voltage outputs (+3.3 V) to power the two photodetectors and the temperature sensor, as well as three analog voltage inputs (0--3.3 V) to receive voltage signals from each component. The analog signals are converted into digital signals with 12-bit digital resolution, which---distributed across the input voltage acceptance range---gives a voltage conversion resolution of 0.81 mV. The signal was transmitted to a computer by means of a micro-USB to USB cable. Once the computer intercepted the data from the Trinket, the information was logged into a CSV file by a Python script running on this computer. Details of this protocol and the associated Python script are provided in the github link at the end of this paper. 

Preliminary quadrature signal analysis was performed in real-time by the Trinket by examining the signals from Photodetectors A and B (refer back to Fig.~\ref{schematic}) as they were acquired, and periodically recomputing the solution to Eq.~(\ref{phaseextract}) in response to the changing photodetector input signals. To accommodate laser intensity variations as well as variations in beam overlap, the microcontroller was additionally tasked with offsetting the raw data sets by their respective center values, as these center values were found to vary.

\section{Results}
\label{sec:results}

\begin{figure*}[tp]
    \centering
    \includegraphics[width = 6.8in]{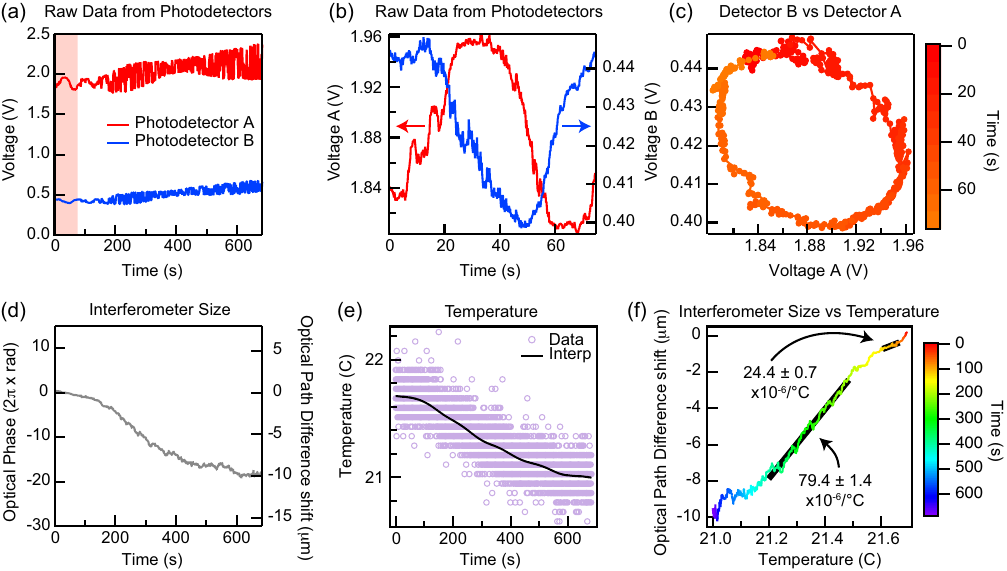}
    \caption{Example data set illustrating the thermal expansion coefficient extraction process. (a) Raw photodetector voltage outputs. (b) A section of data from (a) restricted to the experiment's first 75 seconds (i.e., the region from panel (a) highlighted in red). (c) Lissajous plot of Detector B vs.\ Detector A. (d) Extracted optical phase shift (left axis) and associated optical path difference shift (right axis) of the data depicted in (a). (e) Interferometer temperature vs.\ time as measured with the LM35 sensor (purple circles), and associated smoothing spline interpolation (black line). (f) Interferometer size vs.\ temperature. The slope of this graph can be divided by the overall optical path difference between the interferometer's two arms ($24.0 \pm 0.4$ cm) to yield an estimate of the baseplate's linear expansion coefficient $\alpha$.}
    \label{fig:dataflow}
\end{figure*}

To characterize our device functionality, we adjusted the air conditioner settings in the room where the interferometer was housed, while the interferometer was left running on top of a set of books sitting on a rigid lab desk. Then, we correlated real-time interferometer readings against the LM35 temperature sensor reading to extract the coefficient of thermal expansion of the aluminum plate on which the device was mounted. We compared our results to thermal expansion coefficient values that have been reported in the literature. 

Generally speaking, the thermal expansion properties of materials can be somewhat complicated functions; aluminum expands more rapidly with temperature at room temperature than it does near absolute zero, and water actually contracts when increasing in temperature from solid to liquid state. For small temperature fluctuations and in the absence of phase transitions, however, these functional dependences can be linearized using Taylor series approximations. The standard equation used to define the expansion coefficient under such circumstances is
\begin{equation}\label{expansion relationship}
    \frac{\Delta L}{L} = \alpha \Delta T,  
\end{equation}
where $\Delta L$ is the change in object length, $L$ is the overall object length, $\Delta T$ is the change in temperature, and $\alpha$ is the coefficient of thermal expansion. Solving for $\alpha$ gives 
\begin{equation}\label{coefficient of thermal expansion}
    \alpha =  \frac{\Delta L}{\Delta T} \frac{1}{L}.
\end{equation}
The quantity $\Delta L/L$ is essentially the same as the change in the optical path difference divided by the overall optical path difference between the interferometer's two arms, and so we can see from Eq.~(\ref{coefficient of thermal expansion}) that the thermal expansion coefficient can be extracted by ascertaining the slope of a graph plotting this change in optical path difference as a function of temperature change, and then dividing the result by the overall path length difference, which we measure in our device to be $24.0 \pm 0.4$ cm.

Figure \ref{fig:dataflow} illustrates the data flow of one of our experimental runs. Figure \ref{fig:dataflow}(a) shows the raw data outputs of Photodetectors A and B as a function of time. Figure \ref{fig:dataflow}(b) shows a zoomed-in version of this, corresponding to the first 75 seconds worth of data acquisition, and Fig.~\ref{fig:dataflow}(c) shows a Lissajous figure of the output of Photodetector B vs.\ Photodetector A over this same 75-second time frame. As can be seen, particularly in Fig.~\ref{fig:dataflow}(c), the signals from the two different photodetectors are close to being in quadrature, but not quite perfectly so as evidenced by the fact that the Lissajous figure ellipse exhibits a slight diagonal elongation. This is likely due to a combination of imperfections in the 3D movie glass quarter wave plate (optical retardance is generally speaking expected to be different for different wavelengths and therefore not likely optimized perfectly at 532 nm) and laser beam alignment and wavefront imperfections. 

Figure \ref{fig:dataflow}(d) shows the phase change $\Delta \phi$ (left axis) and change in optical path difference $\Delta (\ell_2-\ell_1') = \Delta \phi/k_0$ (right axis) over the entire trial, calculated by applying Eq.~(\ref{phaseextract}) to the data from Fig.~\ref{fig:dataflow}(a) after subtracting off the center value and normalizing the signal deviations away from this center value to unity. Minimum, maximum, and center values were in their own right calculated in post-processing using a script similar to (but not quite identical with) the real-time phase calculation reported in Sec.~\ref{sec:setup}. Following the phase extraction in $2\pi$-modulo form, the phase was computationally unwrapped, leading to the data that have been ultimately presented.

Figure \ref{fig:dataflow}(e) shows temperature vs.\ time for the full duration of the trial as measured using the LM35. Discretely spaced vertical temperature readings in the panel originate from the finite-granularity of the microcontroller's 12-bit analog-to-digital converter (0.81 mV voltage granularity translates over into a temperature-reading granularity of 0.081$^\circ$ C). These can be averaged away by means of a smoothing spline interpolation as illustrated by the panel's solid black line.

Figure \ref{fig:dataflow}(f) shows the optical path difference shift from Fig.~\ref{fig:dataflow}(d) plotted against the smoothing spline interpolation of the temperature data shown in Fig.~\ref{fig:dataflow}(e). Interestingly, the plot shows that it is generally true that temperature and the interferometer's thermal expansion properties are correlated, which is a fact that is also apparent by examining Figs.~\ref{fig:dataflow}(d) and \ref{fig:dataflow}(e) directly. However, the relationship is actually not a strictly linear one as would have been predicted by Eq.~(\ref{expansion relationship}). There are times when the slope of the plot is gentler (for example, between 75--125 seconds as in the graph's upper right portion), leading to an extracted thermal expansion coefficient of $\alpha = (24.4 \pm 0.7) \times 10^{-6}/^\circ$C. More often, however, the slope is steeper. Between 200--400 seconds, for example (middle of the graph), the extracted thermal expansion coefficient 
turns out to be $\alpha = (79.4 \pm 1.4) \times 10^{-6}/^\circ$C. While the first of these two values is in reasonable agreement with thermal expansion coefficients reported in the literature for aluminum (typically quoted near $23.6 \times 10^{-6}/^\circ$C),\cite{ASMhandbook} the second is clearly not. The results indicate that although the interferometer serves as a useful demonstration piece illustrating the basic functionality of a polarization-based quadrature interferometer, it falls short of being able to be used for more quantitative measurements. Discrepancy origins may include laser Poynting vector stability issues over long periods of time, different parts of the interferometer changing temperature at different rates (although this may ultimately be unlikely given the close proximity of the LM35 to the interferometer base plate), and/or thermally contracting or twisting 3D-printed mounts. We note that the temperature-dependent refractive index of air will have an effect on the interferometer output signal in addition to the aluminum baseplate's physical contraction. For a temperature drop of 0.7$^\circ$ C near a starting temperature of 21.7$^\circ$ C as displayed in Fig.~\ref{fig:dataflow}(e), the overall expected air-induced phase shift is merely 1.90 radians,\cite{toolbox} far below the 120-radian phase shift that is experimentally observed in Fig.~\ref{fig:dataflow}(d).

\section{Conclusions}

The interferometer presented in this paper represents a working device capable of illustrating the qualitative functionality of polarization-based quadrature interferometry basics at a fraction of the cost of the cheapest commercially available alternatives. Components leading to the biggest reduction in these overall costs include a generic green laser pointer, 3D-printed optics mounts, low-cost commercially available polarization optics, home-built photodetectors, and low-cost microcontroller-based analog-to-digital signal conversion.

Looking toward the future, we envision design tweaks that may improve device accuracy with only a marginal increase in cost, potentially leading to quantitatively accurate measurements. Temperature measurements may be able to be improved, for example, by means of better thermal contact established between the temperature sensor and aluminum baseplate and possibly a preamplifier inserted between the sensor output and microcontroller analog-to-digital input. Alternatively, the setup may be modified to utilize an infrared temperature sensor to measure the exact temperature of the aluminum. Stronger adhesives might be applied to the optical mounts in order to better secure them in place. Interferometer application goals could be reoriented to focus on phenomena occurring on faster time scales than temperature fluctuations like vibrational phenomena or turbulence in gasses.

Applications of the interferometer in present and future forms may include use as a classroom demonstration model, and deployment of many devices or device kits in tandem to groups of students taking laboratory optics classes. Beyond this, our hope is that the summary of design elements reported in this work will inspire independent device construction, development, and improvements by readers both inside of academia and beyond.

\begin{acknowledgments}
We thank P.\ T.\ Beyersdorf, H.\ B.\ Wahhab, and M.\ Rojas-Montoya for useful discussions.
This material is based upon work supported by the National Science Foundation under Grant No.\ 2003493.\\
\\
The authors have no conflicts of interest to disclose.\\
\\
Data availability: Data supporting the findings of this study are available upon request from the corresponding author. A selection of data processing scripts and 3D-printer CAD drawings has also been made publicly available on GitHub at \url{https://github.com/Pnguyenkhang/Low-Cost_Quad-Interferometer}.
\end{acknowledgments}


\end{document}